\documentclass[aip,cha,amsmath, amssymb, graphicx, reprint]{revtex4-1}

\usepackage{graphicx}

\draft % marks overfull lines with a black rule on the right

\begin{document}

%\preprint{}

\title{Fractal Dimension Computation From Equal Mass Partitions} %Title of paper



\author{Yui Shiozawa}
\email{yui.shiozawa@tcu.edu}
%\thanks{}
\author{Bruce N. Miller}
\email{b.miller@tcu.edu}
\affiliation{Department of Physics and Astronomy, Texas Christian University, Fort Worth, TX 76129, USA}

\author{Jean-Louis Rouet}
\email{jean-louis.rouet@univ-orleans.fr}
\affiliation{Institut des Sciences de la Terre d'Orl\'{e}ans, France}


\date{29 December 2013}

\begin{abstract}
While the numerical methods which utilizes partitions of equal-size, including the box-counting method, remain the most popular choice for computing the generalized dimension of multifractal sets,  two mass-oriented methods are investigated by applying them to the one-dimensional generalized Cantor set. We show that both mass-oriented methods generate relatively good results for generalized dimensions for important cases where the box-counting method is known to fail. Both the strengths and limitations of the methods are also discussed. 
\end{abstract}
 
\pacs{}% insert suggested PACS numbers in braces on next line

\maketitle %\maketitle must follow title, authors, abstract and \pacs

\begin{quotation}
Fractal sets are characterized by self-similarity, and power laws can be associated with them. Examples of fractals in nature are ubiquitous.  Their discovery led to the extension of the notion of dimension. For  monofractals, the scaling pattern is homogeneous while it varies over the set for multifractals. By introducing the generalized dimension $D_q$, not only a non-integer dimension can be assigned to a set, but also a spectrum of dimensions can be attributed to a single set if the set is a multifractal. In finding the generalized dimensions, the box-counting method has been by far the most popular choice among researchers across various fields. However, it is known that the class of methods which deal with partitions of equal size, including the box-counting method, is ill-suited for computing the generalized dimensions on some domain of $q$. In this paper, two promising methods which utilize mass-oriented partitions, rather than partitions of equal-size, are investigated. 
\end{quotation}


\section{Introduction \label{sec: intro}}

Fractals are the mathematical sets characterized by self-similarity. While the history of the study of fractal goes back as far as the 17\textsuperscript{th} century, \cite{pickover2009} the concept was popularized by Mandelbrot in 1970s \cite{mandelbrot82} and is now applied to many fields from cosmology, \cite{pietronero1987} and chemistry, \cite{Hasal94} to economics. \cite{serletis2004} The fact that fractals can be found virtually everywhere suggests that there is an underlying mathematical principle. From a geometrical perspective, a given set is self-similar when it is similar part of itself. A moment of thought convinces us that, to achieve this condition, a self-similar set needs to possess an infinite nesting structures. Due to this self-similarity, fractals may be preserved after appropriate magnification and translation. Therefore, power laws arise naturally in the study of fractals as the power law is the only differentiable function that does not change its form under a scale transformation. To be precise, if for some differentiable function $f$ which satisfies $f(bx)=g(b)f(x)$ for all $b>0$ for some function $g$,  the function $f$ must be a power law. Since $x \rightarrow bx$ is a scale transformation, the function $f$ is said to be preserved up to a constant under a scale transformation. Accordingly, various power laws can be derived from fractal sets and it is the exponents of these power laws that the dimensions of the fractal set are associated with. 

Traditionally, the dimension of a given set indicates the number of independent variables required to specify the element within the set and so can take only integer values. However, if we want to associate ``size" with fractals such as the famous Koch snowflake, \cite{Koch1904} we need to extend our notion of dimension as well as of measure. The Koch snowflake is nowhere differentiable and consists of a perimeter with infinite length enclosing a finite area. Intuitively, the dimension of the set should be bigger than a finite interval and smaller than a finite area. Indeed, we can define the fractal dimension in a way that the Koch snowflake has the dimension of $\log3/\log4=1.261...$. In this example, the fractal dimension is smaller than the topological dimension in which it is embedded. Note that the fractal dimension can be non-integer. Here, only the single dimension is associated with the set and so the Koch snowflake is said to be monofractal. Monofractals are a type of fractal for which the associated power laws are homogeneous within the whole set. If more than one scaling law, and therefore the corresponding exponents, are required, the set is said to be multifractal. Accordingly, a single dimension cannot fully capture the dimensionality of multifractal sets. To resolve this issue, the generalized dimension $D_q$ was introduced by R\'{e}nyi. \cite{renyi1960} The index $q$ can take any real number and therefore, a spectrum of dimension can now be attributed to a given set. For a monofractal, the generalized dimension $D_q$ is constant for any $q$. In this formulation, more familiar fractal dimensions such as the box-counting ($D_0$), the information dimension ($D_1$) \cite{farmer83} and the correlation dimension ($D_2$) \cite{Itamar83} are said to be special cases of the generalized dimensions. 
	However, since few fractals can be characterized analytically, the search for effective numerical methods is inevitable. Thus far, the box-counting method has been the most popular among researchers despite its difficulty to accurately compute the generalized dimension on some domains of $q$. The difficulty is rooted in the fact that numerical methods are required to deal with a finite representation of true fractal sets. Therefore, the sampling process from a theoretical set needs to be carefully handled. 
	
	In this work, two promising numerical methods for obtaining generalized fractal dimensions are examined. One of the methods utilizes the probability distribution of the nearest neighbor distances among randomly chosen points within a given set. \cite{Badii85} The other method involves the collection of distances of the $k$\textsuperscript{th} nearest neighbor as $k$ increases. \cite{Water88} They can be applied, in principle, to any set as long as a sufficient number of sample points can be taken from the set. Unlike the box-counting method, which employs a partition composed of equal-sized cells, the two methods examined in this paper employ mass-oriented partitions. The nearest neighbor method utilizes partitions composed of equal-mass cells while the k-neighbor method uses partitions composed of cells with cumulative mass. These alternative approaches enable one to compute the generalized dimension on the domain where the box-counting method encountered difficulty. Another advantage of these methods is their ability to generate a spectrum of generalized dimensions almost simultaneously, and therefore they are particularly suited for the analysis of multifractals. 
	
	This work was originally motivated by the emergence of fractal patterns on the one-dimensional universe model. \cite{miller2010ewald, miller2010cosmology} Thus, our focus is on one-dimensional sets although the numerical methods used in this paper can be applied to higher dimensional spaces. The analysis of fractal dimension should give us some insight into the fractal structures which arise in many chaotic systems. In particular, we applied the methods to the generalized Cantor set. The generalized dimensions of the generalized Cantor set can be readily derived analytically, thus enabling the accuracy of the numerical methods to be verified. We sampled points from the finite representation of the generalized Cantor set according to the weight assigned to each interval. In general, numerical methods need to deal with finite samples which often gives rise to technical difficulties. No finite sample is a true fractal set, and therefore, the statistical data extracted from a finite sample may not accurately reflect the property of the original set one wishes to study. It is worth noting that simply increasing the number of sample points from an available data set can partially overcome the difficulties associated with numerical methods. While a true mathematical fractal is characterized be an infinite nesting structure, ``fractals" found in nature have a limited hierarchal structures and the range where a power law is observed is finite. Accordingly, when employing a numerical method, one is required to determine the applicability of the method in relation to a finite sampling process. The generalized Cantor set is an ideal set in that the degree of hierarchy can be readily controlled.  
	It turns out that the nearest neighbor method suffers from the presence of singularities on a certain domain of $q$ in the generalized dimensions, and therefore the range on which the method provides a reliable result is restricted. Nevertheless, for the computation of the box-counting dimension ($q=0$) as well as $D_q$ for $q$ near 0, both methods managed to generate results which agree well with the theoretical values within a reasonable amount of computational time.
		
	The paper is organized as follows: In section \ref{sec: def}, the important definitions and notations are stated. In section \ref{sec: numerical}, we explain the nearest neighbor method and the $k$-neighbor method in depth. In section \ref{sec: implementation}, we discuss some of the issues particular to numerical simulations. Section \ref{sec: results} includes an overview of our results and various raw data obtained using the aforementioned methods. Mathematical methods are employed to analyze the results in section \ref{sec: analysis}. In section \ref{sec: conclusion}, a summary and conclusions are provided. 

\section{Definitions \label{sec: def}}

\subsection{Generalized Cantor Set}
The Cantor set is one of the most iconic fractals and readily generalized to a multifractal set. Accordingly, we use the generalized Cantor set as our seminal test set to which the numerical methods are applied. It is constructed in the following way: It starts with a interval of unit length. Then take out the middle part of the interval in such a way that the remaining interval on the left has a length of $l_0$ and on the right $l_1$. Moreover, a weight is assigned to each interval, namely $p_0$ or $p_1$, such that $p_0+ p_1=1$ The same procedure is applied to each of the two remaining intervals which then results in four intervals with lengths, starting from the left, $l_0^2$, $l_0l_1$, $l_1l_0$, $l_1^2$ and weights $p_0^2$, $p_0p_1$, $p_1p_2$, $p_1^2$. In general, after $m$ such iterations, $2^m$ intervals with various factors remain. A generalized Cantor set is what remains after taking $m\to\infty$. Particularly, a standard uniform Cantor set is obtained for $l_0=l_1=\frac{1}{3},  p_0=p_1=\frac{1}{2}$. Another special case, referred to as the multiplicative binomial process, or MBP, is defined by  $l_0=l_1=\frac{1}{2}$ with arbitrary weights. \cite{Altham78} 

Note that, unless $m= \infty $, the set is not a true Cantor set. For finite $m$, the set will be referred to as the finite representation of the Cantor set with hierarchy degree $m$. Now, on the $m$\textsuperscript{th} degree, the weight assigned to each interval is given by $p_k^{(m)}=p_0^{m-k}p_1^{k}$.  The index $k$ runs from $0$ to $m$ and depends on the location of the associated interval. Similary, we can denote the length of each segment on the $m$\textsuperscript{th} level by $l_k^{(m)} = l_0^{m-k}l_1^{k}$. Then there exists an $\alpha_k \in \mathbb{R}$ such that $p_k^{(m)}=(l_k^{(m)})^{\alpha_k}$. In general, $\alpha_k$ depends on $k$ unless $p_0=p_1$ and $l_0=l_1$. Such $\alpha_k$ is called the local dimension or singularity. Therefore, the uniform Cantor set has single value for $\alpha_k$ and is said to be monofractal. Otherwise, generalized Cantor sets are multifractal, meaning that the local dimension varies from place to place within a set. If $N_{\alpha_k}$ denotes the number of segments with the local dimension $\alpha_k$, we define $f(\alpha_k)$ such that it satisfies the following relation:
\begin{equation}
 N_{\alpha_k }= (l_k^{m})^{-f({\alpha_k})}
\end{equation}in
$f(\alpha)$ is called the spectrum of scaling indices \cite{halsey86} and is related to the R\'{e}nyi Dimension that is discussed in the next section.

\subsection{R\'{e}nyi Dimension}
As mentioned in the introduction, the traditional notion of dimension can be extended to generate a spectrum of dimensions for a given set. While a single characteristic dimension is associated with monofractals, a spectrum of dimensions is required to reflect the properties of multifractals. Suppose $C={\{U_i \}}$ is a cover of a set $A\subset \mathbb{R}^n$. Let $n_i$ denotes the number of points in $U_i$ among $n$ randomly chosen points from $A$. Then $p_i$ is associated with $U_i$ for each $i$ by $p_i=\lim_{n \to \infty}{\frac{n_i}{n}}.$ For any real number $q\neq1$, the generalized dimension $D_q$ for a set $A$ is given by \cite{Hentschel83}

\begin{equation}
D_q=-\frac{1}{1-q}\lim_{\epsilon \to 0} \frac{\ln \sum_{i=1}^{N(\epsilon)} p_i^q}{\ln \epsilon} \label{eq: Renyi} \end{equation}
where $N(\epsilon)$ is the number of sets with diameter $d(U_i)=\epsilon$ required to cover the set $A$. For $q=1$, the limiting case where $q\to1$ is used. The topological dimension can be recovered when applied to traditional geometries and in particular, $1$ for a line interval. The generalized dimension is also known as the R\'{e}nyi Dimension, named after a Hungarian mathematician, Alfr�d R\'{e}nyi as it can be formulated using the R\'{e}nyi entropy $K_q$,
\begin{equation}
 K_q = \frac{ \ln \sum_{i=1}^{N( \epsilon)}{p_i^{q}}} {1-q}. \label{eq: entropy}
\end{equation}
Eq. (\ref{eq: entropy}) can be regarded as the generalized form of Shannon's entropy.
In fact, in the limit of $q \to 1$, the R\'{e}nyi entropy $K_q$ is reduced to the familiar equation:
\begin{equation}
K_1=-\sum^{N}_{i=1}p_i \ln p_i.
\end{equation}
Using the R\'{e}nyi entropy $K_q$, the generalized dimension can be formulated as:
\begin{equation}
D_q = -\lim_{\epsilon \to 0} \frac{K_q(\epsilon) } {\ln{\epsilon}}
\end{equation}
Note that when $q=0$, the R\'{e}nyi dimensions coincides with the box-counting dimension $D_0$.
\begin{equation}
D_0=- \lim_{\epsilon \to 0} \frac{ \ln N(\epsilon)}{ \ln \epsilon}
\end{equation}
In other words, for sufficiently small $\epsilon$, the following relation is satisfied:
\begin{equation}
N(\epsilon) \simeq \epsilon^{-D_0}
\end{equation}
The equation above is an example of the power law relations that can be derived from a given set. Note that the box-counting dimension has the opposite sign of the exponent. 

In the case of the $m$\textsuperscript{th} finite representation of the generalized Cantor set, the natural cover would be the broken intervals themselves and so the weight of each interval $p_k^{(m)}$ may be used for $p$ in Eq. (\ref{eq: Renyi}). Then it can be readily shown that for the uniform Cantor set with $l_0=l_1$ and $p_0=p_1$, we have 
\begin{equation}
D_q=\frac{\ln2}{\ln3} \label{eq: UniDimension}
\end{equation}
for all $q$. Therefore, the R\'{e}nyi dimensions of the uniform Cantor set are $q$-independent and hence a monofractal. On the other hand, applied to the MBP, it can be shown that  \cite{Hentschel83}
\begin{equation}
D_q=\frac{1}{q-1}\frac{\ln (p_1^q+p_2^q) }{ \ln 2 } \label{eq: MBPDimension}
\end{equation}
where $p_1$ ($p_2$) is the weight of the left (right) interval and $l=l_1=l_2=\frac{1}{2}$ the length of the segments at the first iteration. Thus, the MBP is a multifractal set. There is no explicit formula for $D_q$ when $l_1\neq l_2$, but the dimension $D_q$ can be found from an implicit relationship that employs the spectrum of scaling indices $f(\alpha)$ and
 the Legendre transform. \cite{halsey86} For a general set, it is often difficult, if not impossible, to find appropriate covers. Thus methods which permit numerical simulations should be sought. 

\section{Numerical Methods \label{sec: numerical}}
In this section, three numerical methods for computing the R\'{e}nyi Dimensions are discussed.

\subsection{Box-Counting Method}
This method is probably the most well-known and is closely related to the original definition of the R\'{e}nyi Dimensions. There are a few slightly different versions under the name of the box-counting methods, using ``spheres" instead of ``boxes" for example, \cite{caswell86} but the underlying ideas are similar: generally, the number of cells required to cover the points in a given set, $n$, changes as the size of the partitions $\epsilon$ changes. The scaling relation can be extracted for a fractal set as the size of the partitions decreases, namely,
\begin{equation} 
D = - \lim_{\epsilon \to 0} \frac{\ln n(\epsilon)}{\ln \epsilon}
\end{equation}
Due to the simplicity of the method, it is widely used among researchers. However, it has been pointed out by many that this method and, more generally, methods that involve partitions of the same size such as the Correlation method, do not work well for $q < 1$. \cite{Badii88} A heuristic explanation is given below to understand this result.  In Eq. (\ref{eq: Renyi}), we can see that if $q>1$, the contribution from relatively large $p_i$ is emphasized, and if $q<1$, the contribution form relatively small $p_i$ plays the dominant role. The larger the value of $  \lvert q \rvert $, the greater the effective discrimination. Therefore, the fact that the method does not produce a good result for $D_q$ with $q<1$ means that the sparse regions of the set are not well-represented in the finite representation of the Cantor set. Since a true fractal possesses an infinite number of points or elements, any finite set may not be large enough to represent the true Cantor set in relatively sparse regions. In some instances, a finite representation of a fractal may be thought of as a subset of a corresponding fractal as in the H\'{e}non map. \cite{Henon76} As the size of the cells diminishes, the truncated finite sample no longer statistically represents the sparse regions of a true fractal. Under the same condition, the dense regions are affected less by the finite size effect. Since numerical methods always have to deal with a finite sample, different methods need to be considered to find an accurate result for $q<1$.

\subsection{Nearest Neighbor Method}

\begin{figure}[h]
 \includegraphics[scale=0.65]{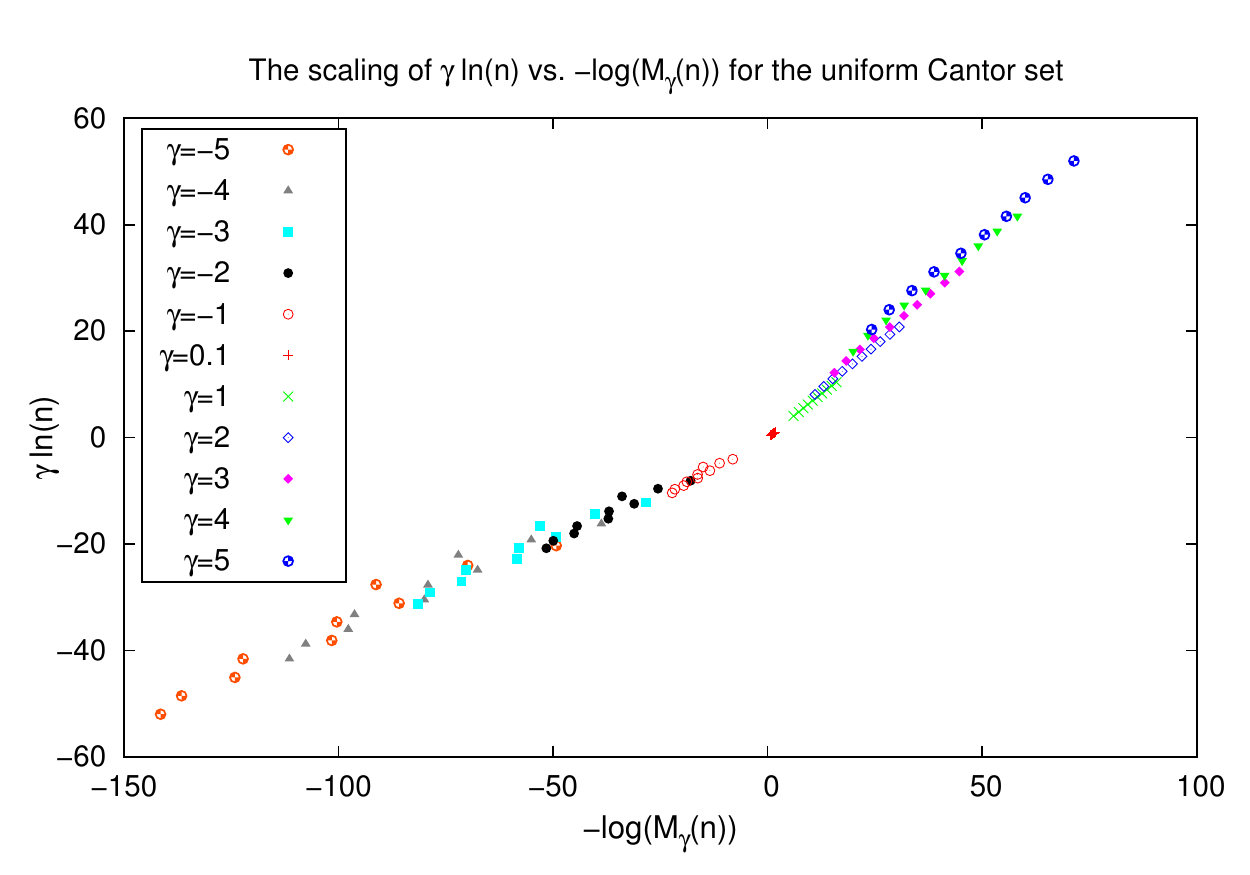}
 \caption{ For the uniform Cantor Set, $\gamma \ln(n)$ vs. $-\ln (M_{\gamma}(n))$ is plotted for each $\gamma$ as $n$ is increased. According to Eq. \ref{eq: DF}, the slope converges to $D(\gamma)$. The corresponding result for $D(\gamma)$ is shown in Fig.~\ref{fig: D_q_UCS} \label{fig: Scaling_Slope}}
 \end{figure}

\begin{figure}[h]
 \includegraphics[scale=0.65]{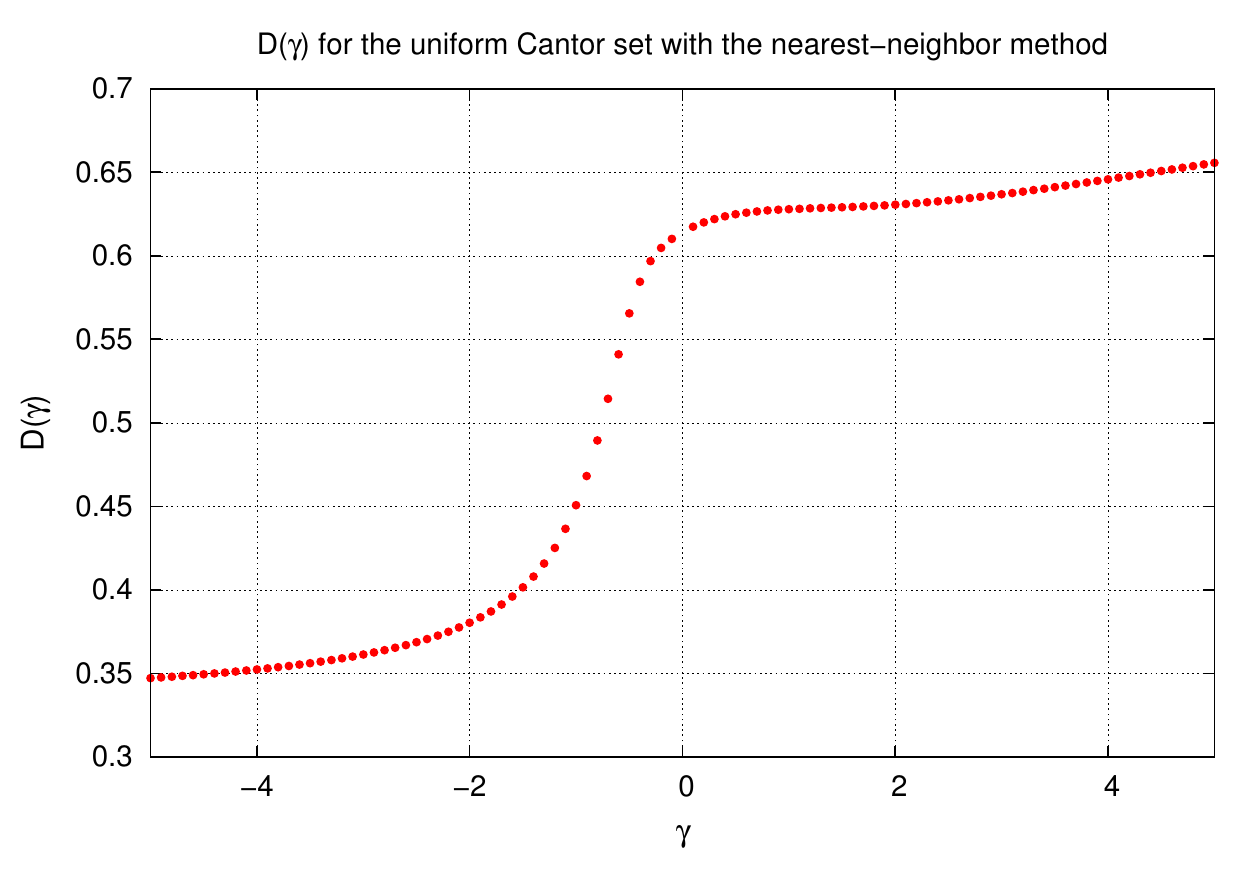}
 \caption{In this graph, the Dimension Function $D(\gamma)$ for the Uniform Cantor set was computed as the slope of the best-fit line to the corresponding data set which is partially plotted in Fig. ~\ref{fig: Scaling_Slope}. $D(\gamma)$ diverges from the analytical result which is $\log2/\log3$ for negative $\gamma$.   \label{fig: D_q_UCS}}
 \end{figure}

\begin{figure}[h]
 \includegraphics[scale=0.65]{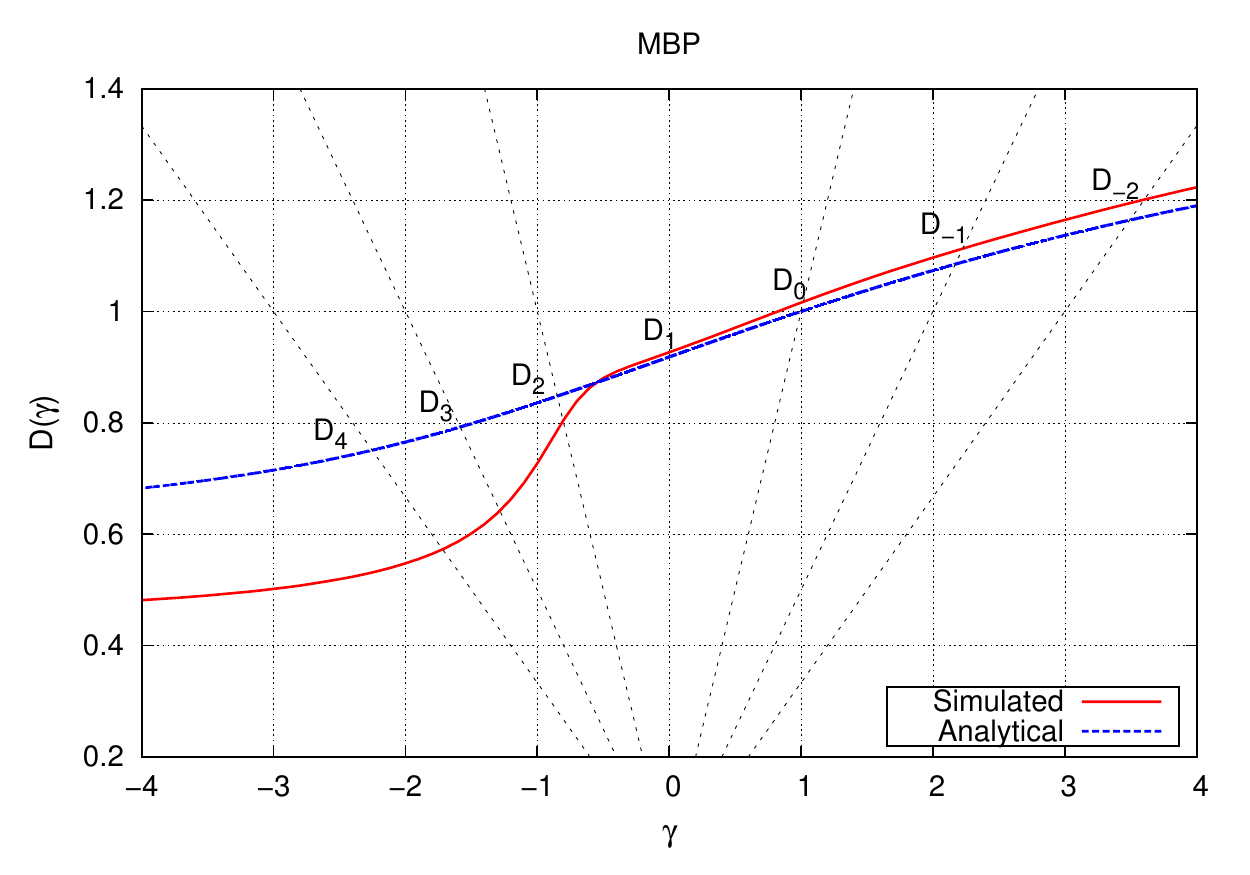}
 \caption{The solid curve is the simulated result of the Dimension Function for MBP. Note how $D_q$ can be obtained by locating the corresponding intersections. For example, the box-counting dimension $D_0$ can be found at the intersection of $D(\gamma)$ and $y=\gamma$. \label{fig: Dq}}
 \end{figure}

The approach called the ``nearest neighbor method'' was first introduced by Badii and Politi.\cite{Badii85} This method  is essentially based on their observation that \begin{equation} \label{eq: NN assumption}
<\delta> \ \sim n^{- \frac{1}{D_0}}
\end{equation} where $<\delta>$ denotes the mean distance from each point to its nearest neighbor among $n$ randomly chosen points from a given test set and, as discussed earlier, The value $D_0$ is just the box-counting dimension. By naturally extending the premise, the Dimension Function $D(\gamma)$ can be computed by using the moments of order $\gamma$ of the distribution function $P(\delta, n)$ generated by an ensemble of $n$ randomly chosen points:
\begin{equation}
<\delta^{\gamma}> \equiv M_\gamma(n) \equiv \int_0^\infty \delta^\gamma P(\delta , n)d\delta = Kn^{-\frac{\gamma}{D(\gamma)}} \label{eq: integral}
\end{equation}
where $K$ is some function of $n$ and $\gamma$ which asymptotically remains bounded as $n$ becomes large. Here, the meaning of $\gamma$ should be clear; the dense region of a given set generates smaller values of $\delta$, the distance to the nearest neighbor, and vice versa. The proof of a more general relation is provided by van de Walter and Schram. \cite{Water88} It follows that the Dimension Function $D(\gamma)$ can be obtained by:
\begin{equation}
D(\gamma) =-\lim_{n \to \infty} \frac{\gamma \ln n}{\ln M_\gamma(n)}  \label{eq: DF}
\end{equation} 
The function $K$ generally depends on $n$ and $\gamma$ but $K$ should be, by definition, irrelevant in the limiting case as in Eq. (\ref{eq: DF}). In numerical analysis, the value of $K(n,\gamma)$ does affect the numerical result as $n$ is finite. The Dimension Function $D(\gamma)$ can be thought of as an alternative generalized dimension and is related to the R\'{e}nyi Dimension by: \cite{Badii85}
\begin{equation}
D[\gamma = (1-q)D_q] = D_q
\end{equation}
As the equation suggests, once $D(\gamma)$ is obtained, the generalized dimension $D_q$ can be found as the intersection of $D(\gamma)$ and the straight line with slope $(1-q)^{-1}$ which passes through the origin as illustrated in Fig. ~\ref{fig: Dq}.

For most cases, the generalized dimension $D_q$ is uniquely determined from $D(\gamma)$. Note that a larger $q$ does not correspond to a larger $\gamma$ due to the negative sign in the equation. Therefore the index $\gamma$ plays a similar role as $q$ in that it  discriminates the range of density of a given set that most strongly contributes to $D(\gamma)$. In simulations, the Dimension Function $D(\gamma)$ is obtained using Eq. (\ref{eq: DF}). The formula can, in principle, be applied to sets with any topological dimension. In the case of a one-dimensional set, sample points are prepared in a way that $\delta$ is bounded from above by 1. Therefore, the integral in Eq. (\ref{eq: integral}) can be taken from 0 to 1. Unlike the box-counting method, this algorithm does not make use of partitions of the same size but, rather, of the same ``mass" for it can be considered that each element of the partition contains two points, namely a reference point and its nearest neighbor. Badii and Politi used a slightly improved version of the method which uses partitions containing three or four points to smooth out local statistical anomalies. \cite{Badii85} Broggi used partitions containing up to 300 points for systems of large dimensionality. \cite{broggi1988evaluation}

\subsection{$k$-Neighbor Method}
Another method called ``k-neighbor" is similar to the nearest neighbor method in that its partitions are taken according to the number of masses inside. However, instead of fixing the number of masses as in the case of the nearest neighbor method, the k-neighbor method incorporates a partition of cumulative mass. In fact, the nearest neighbor method is a special case of the $k$-neighbor method with $k=1$. By not limiting to $k=1$, the scaling property is obtained through the global structure of a given set, and thus the method is less sensitive to local statistical anomalies which often arise in a finite sample set.  A similar global approach was introduced by T{\'e}l et al. \cite{tel1989} using elements of different size, rather than different mass, and some literature misleadingly refers to it as the ``cumulative mass" method. \cite{lopes2009} The k-neighbor method records the distance $\delta(k,n)$ from a reference point to the $k$\textsuperscript{th} neighbor point among $n-1$ randomly chosen points from a given set. van de Water and Schram formulated a technique for evaluating $D(\gamma)$ from the average of $\delta(k,n)^{\gamma}$ by using the local dimension introduced in Section \ref{sec: def}. \cite{Water88} The average of $\delta(k,n)^{\gamma}$ is defined as follows: 
\begin{equation}
\Delta^{(\gamma)} (k,n) = \frac{1}{n} \sum_{j=1}^{n}\delta_j^\gamma(k,n). \label{eq: k-1}
\end{equation} 
where $\delta_{j}(k, n)$ represents the $k$\textsuperscript{th} neighbor distance from $j$\textsuperscript{th} reference point when $n$ points are randomly chosen from a test set. Here, all $n$ sample points are used as reference points. When $n$ is large, it can be shown that 
\begin{equation}
\left<\Delta^{\gamma}(k,n) \right>^{1/\gamma}  \cong n^{-1/D(\gamma)} \left[\alpha D(\gamma) \frac{\Gamma(k+\gamma/D(\gamma))}{\Gamma(k)} \right]^{1/\gamma} \label{eq: k-3}
\end{equation}
where $\alpha$ is some constant independent of $\gamma$.  Note that the average of $\delta^{\gamma}_{j} $ from a single set is used in Eq. (\ref{eq: k-1}) whereas the derivation of Eq. (\ref{eq: k-3}) is based on the ensemble probability. For large $k$, a simple approximate relation can be obtained: 
\begin{equation}
\left[ \Delta^{(\gamma)} (k,n) \right] ^{1/\gamma} \cong n^{-1/D(\gamma)} k^{1/D(\gamma)}G(k,\gamma) \label{eq: k-2}
\end{equation} where $G(k,\gamma)$ is a correction function close to unity. According to Eq. (\ref{eq: k-2}), the Dimension Function $D(\gamma)$ can, in principle, be obtained from the slope of the best-fit straight line in the log-log plot with either a fixed $n$ or $k$. When $k=1$, the equation is reduced to the key relation in Eq. (\ref{eq: integral}) for the nearest neighbor method. With the $k$-neighbor method, we used a fixed value of $n$. The correction function $G(k,\gamma)$ generally exhibits a periodic pattern as a direct consequence of the self-similarity of fractals as seen in Fig. ~\ref{fig: K}. By fixing $n$ instead of $k$, we can extract a global property of a given set, which makes the $k$-neighbor method less sensitive to local anomalies which often arises from a finite sampling process. 

\section{Numerical Implementation \label{sec: implementation}}
When dealing with a fractal set numerically, one needs to confine oneself to a finite sample. For a Cantor-like set, the number of iterations $m$ needs to be finite. The hierarchy degree $m$ should be chosen in a way that two points in neighboring intervals of the set are distinguishable within the precision of a given numerical environment. In our experiment,  $m=30$ is typically used and therefore we assume a finite representation of the Cantor set which consists of $2^{30}$ intervals. Generally, the larger the number of reference points $n$ is used, the more accurate the result would be obtained, but $n$ can only be increased by correspondingly increasing the amount of computation time but, as we discuss below, there is another limitation on $n$ as well. 

When the number of sample points exceeds the number of broken intervals, the expected probability distribution does not produce a desirable result since the distribution within an interval is nothing but that of a line interval. Therefore, the scaling property needs to be obtained for $n$ sufficiently smaller than $2^{m}$ but large enough to accurately reflect a given fractal set. Each of the $n$ points is randomly assigned to a particular one among $2^{m}$ intervals. The position of the point is then randomly chosen within the window of the chosen interval. Therefore, in our model, most of sample points are taken from the points which are not in the real Cantor Set. However, in principle, we can always set the upper limit to the distance between the sample points and closest points in a true set by taking  $m$ sufficiently large. Choosing a particular interval randomly among $2^{30}$ intervals amounts to randomly generating 30 binary digits. This can be seen by assigning 0 to the left interval and 1 to the right interval on each level of the Cantor set. For the uniform Cantor set, the probability of generating 0 and 1 is exactly half. For the generalized Cantor set, the corresponding weight factors are introduced. 

The Mersenne Twister Pseudo-Random Number Generator \cite{Matsumoto98} for c++ was our primary choice for obtaining random numbers. The built-in c++ random number generator was also used. No idiosyncratic behavior from the particular choice of random number generator was observed. Due to the limitation of the size of $n$, an ensemble average must be employed in order to achieve higher accuracy rather than increasing $n$. The number of members of the ensemble required to stabilize the result depends on the range of $\gamma$. See section \ref{subsec: range} for details. 

\section{Results \label{sec: results}}
Generally, with a small amount of computational time, both of the methods in the fixed-mass class give good indications of the R\'{e}nyi Dimension in the vicinity of the box-counting dimension $(q=0)$ on various generalized Cantor Sets. This is a major advantage over the box-counting method if one seeks to find the box-counting dimension. Around the box-counting dimension, the nearest neighbor method yields a result closest to the analytical solutions. However, as $\gamma$ goes away from it, the k-neighbor method produces more accurate results. Therefore, at this point, no single method seems reliable enough for an extended domain $q$ of the generalized dimension. However, the combination of the aforementioned methods reveals the essential features of a given set such as whether it is a monofractal or multifractal. For a multifractal set, how the dimension changes over the domain $q$ is a key property. The k-neighbor seems to be the best method to start with as it can provide the estimate of the generalized dimension over an extended region, albeit not too accurately. To obtain the dimension to a higher accuracy for a particular $q$ or $\gamma$, the box-counting or the nearest neighbor method may be used. For $q>1$, the box-counting method should be employed and for $q<1$, the nearest neighbor, provided that $q$ is not a very large negative number. Therefore, if possible, the results obtained from these methods should be compared and examined to see if they are consistent within the uncertainty of each method. 

\subsection{Nearest Neighbor Method}

\begin{figure} [h]
\includegraphics[scale=0.65]{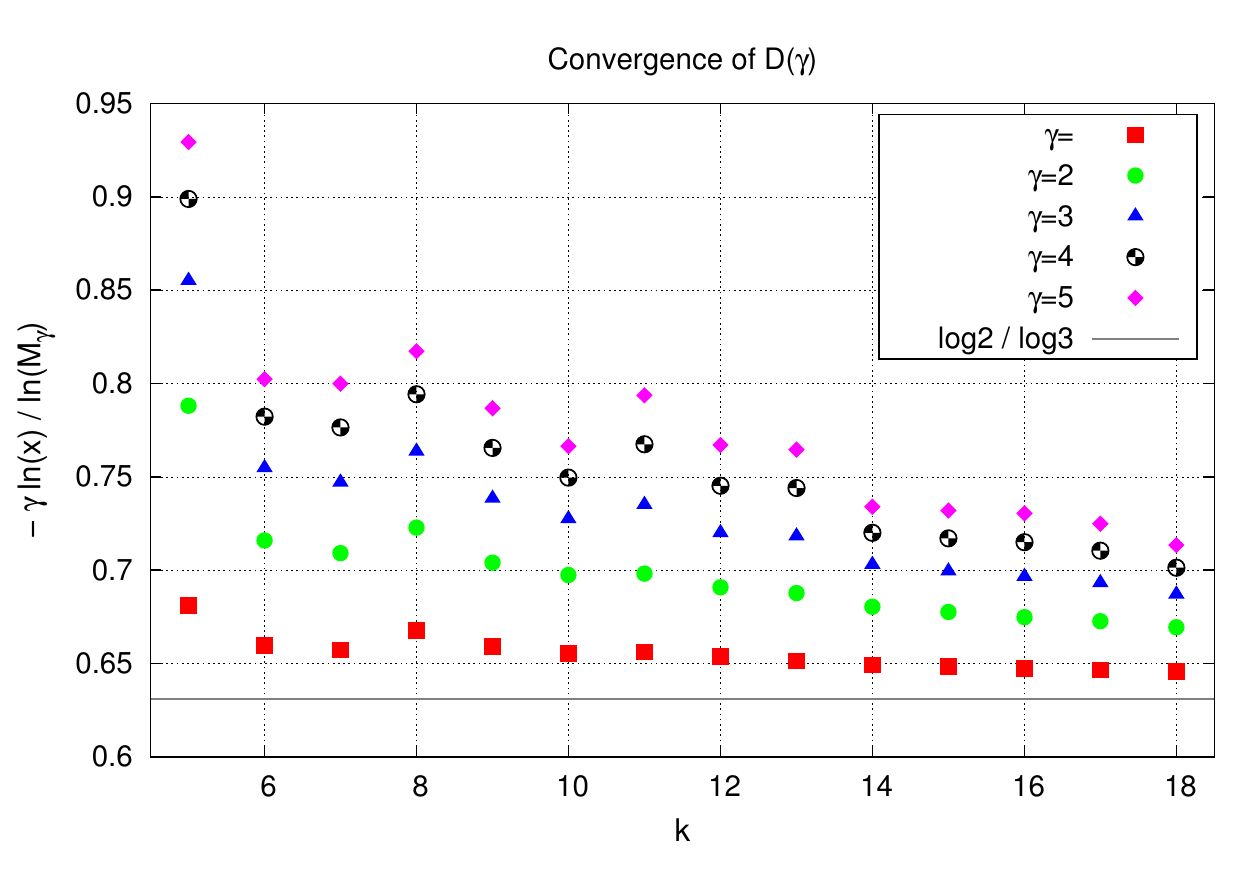}
\caption{\small \sl This figure shows how increasing $n=2^{k}$ affects the value of $- \gamma \ln{n}/\ln{M_{\gamma}}$. The plot was generated for the uniform Cantor set. The analytical value for $D(\gamma)$ for all $\gamma$ is $\log2/\log3=0.630...$ which corresponds to the horizontal line in the plot. \label{fig: convergence}}  
\end{figure}  

\begin{figure} 
\includegraphics[scale=0.65]{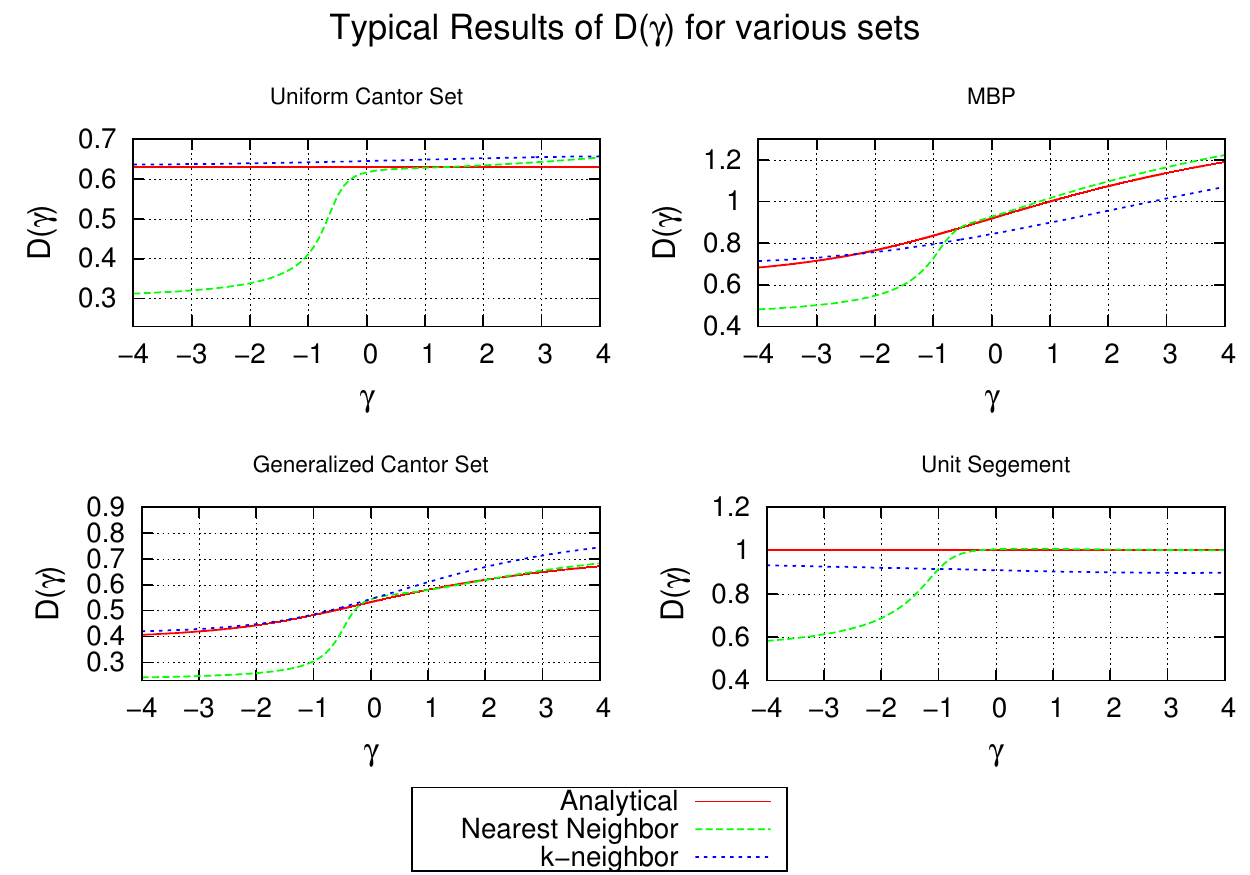}
\caption{\small \sl These plots show the typical results of $D(\gamma)$ for the nearest neighbor method and the $k$-neighbor method applied to four different sets and the corresponding analytical values. ``Unit Segment" here means a interval of unit length and can be thought as the $0$\textsuperscript{th} finite representation of the Cantor set. For negative $\gamma$, numerical results persistently deviate from the analytical results considerably for the nearest neighbor method. While the $k$-neighbor method works relatively well for all $\gamma$, the outcome may not be as accurate as the nearest neighbor method for small positive $\gamma$. \label{fig: Typical}}  
\end{figure}

In the nearest neighbor method, the Dimension Function $D(\gamma)$ was extracted from Eq. \ref{eq: DF}. In Eq. \ref{eq: DF}, the right hand side reads $-\frac{\gamma \ln n}{\ln M_\gamma(n)}$ before taking the limit. To investigate how it approaches to the limit, $\ln{n}/\ln{M_1}$ versus $\ln{n}$ for the uniform Cantor set was plotted in Fig. ~\ref{fig: convergence}. The points in the plot indicates how $- \gamma \ln{n}/\ln{M_{1}}$ seemingly approaches the theoretical limit of $\ln{2}/\ln{3}=0.63...$ as $\ln(n)$ increases in the case of uniform Cantor set. However, it can be seen that the convergence rate is rather slow. Given that $m$ is large enough, increasing $n$ can almost always guarantee a higher accuracy around the box-counting dimension. However, since the convergence rate is rather slow, determining the limit is not a trivial task. For $\gamma=1$, the number of sample points $n=2^{9}=512$ was required to obtain the result within 5\% accuracy and $n=2^{17}$ to obtain the result within 3\%. For quick simulations, we typically used $n=2^{16}$ and $10$ ensembles. In general, we employed the linear regression technique and obtained the limit from the slope of the appropriate log-log plot. While the overall qualitative features of the Dimension Function such as the non-decreasing property are properly reflected on the domain where $\gamma$ is positive, the deviations and the fluctuations around $\gamma=-1$ seem sudden and uncontrolled. The difficulty of obtaining a sensible result for $\gamma<-1$ seems persistent throughout the set we have tested. 
In Fig. ~\ref{fig: Typical}, the results for various generalized Cantor Sets are shown; the domain of $\gamma$ on which the simulated $D(\gamma)$ agrees well with the analytical results is between 0 and 2. For a multifractal, as $\gamma$ increases, the numerical results start to diverge from the analytical result as well. 

\subsection{$k$-Neighbor Method}
\begin{figure}  [h]
\begin{center}  
\includegraphics[scale=0.65]{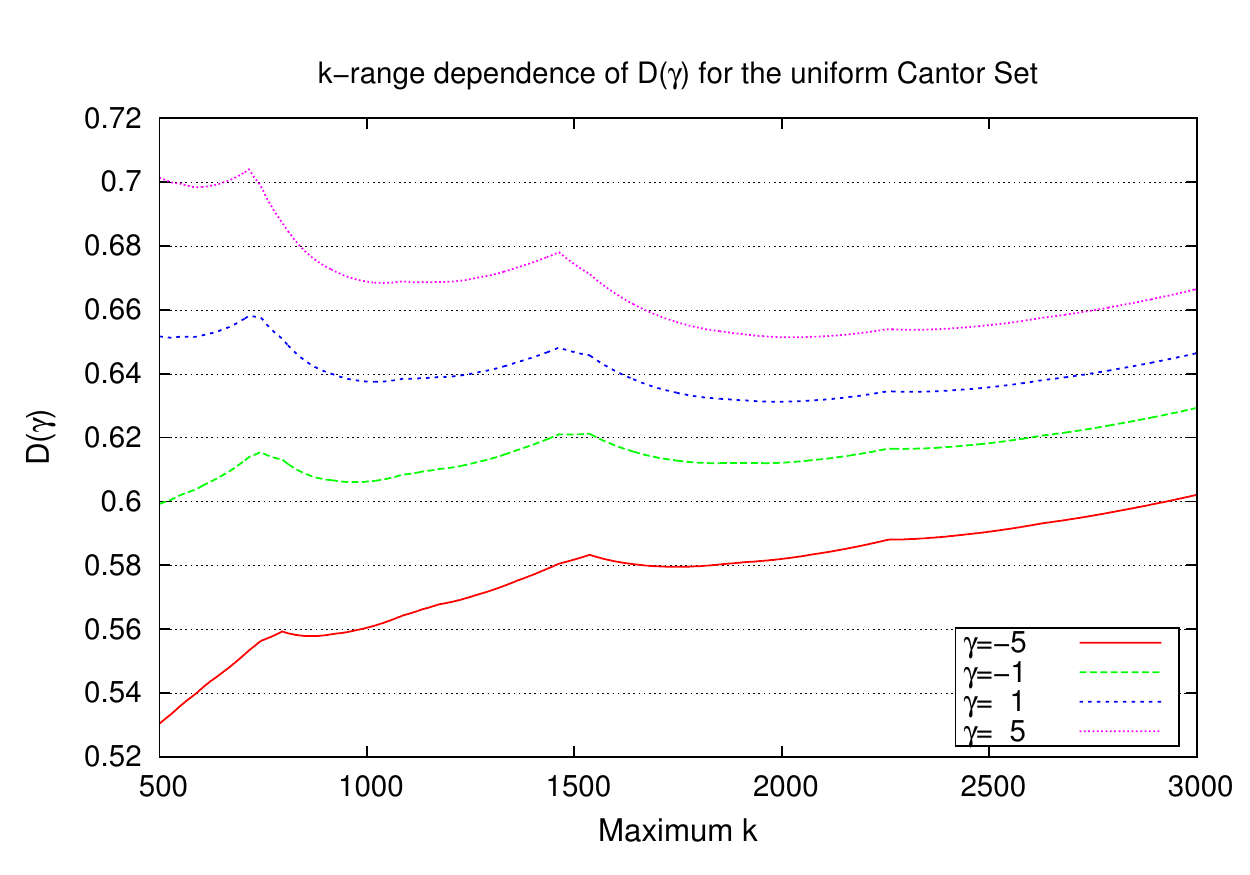}
\caption{\small \sl This plot shows how $D(\gamma)$ differs when a different range is used to extract the slope in the k-neighbor method. For the uniform Cantor set, increasing the upper bound of $k$ generally seems to produce better results. However, this is not a general result. \label{fig: slope}}  
\end{center}  
\end{figure}

Unlike the nearest neighbor method, where the choice of $n$ is often limited by an available finite sample and computational time, the $k$-neighbor method can utilize a larger data set from which the slope is extracted to estimate $D(\gamma)$. As we can see, the fine structure is clearly observed in a log-log plot which injects arbitrariness in a slope-fitting process. This point is covered in detail in section \ref{sec: analysis}. For a fixed value of $n$, $D(\gamma)$ or, to be precise, the corresponding $1/D(\gamma)$ in Eq. (\ref{eq: k-2}), is taken as the slope of $\log{\delta^{\gamma}(k,n)}$ versus $\log{k/n}$. As shown in Fig. ~\ref{fig: K} , the obtained $\delta^{\gamma}(k,n)$ exhibits a periodic pattern, so all approaches to obtain the slope seem to inject ambiguity. We have used the standard linear regression technique  \cite{Press2007} using sample points equally spaced in the logarithmic scale of $k$ rather than in the $k$ scale. Another consideration is that the slope, and therefore, the result for $D(\gamma)$ depends on the range to which the linear regression is applied. It turns out that the best range seems to differ for different $\gamma$ as shown in Fig. ~\ref{fig: slope}. The plot shows how $D(\gamma)$ varies when increasing the upper bound of the slope range when applied to the uniform Cantor set with the analytical dimension of $\log{2}/\log{3}=0.63... $ for all $\gamma$.

\begin{figure}[h]
\includegraphics[scale=0.65]{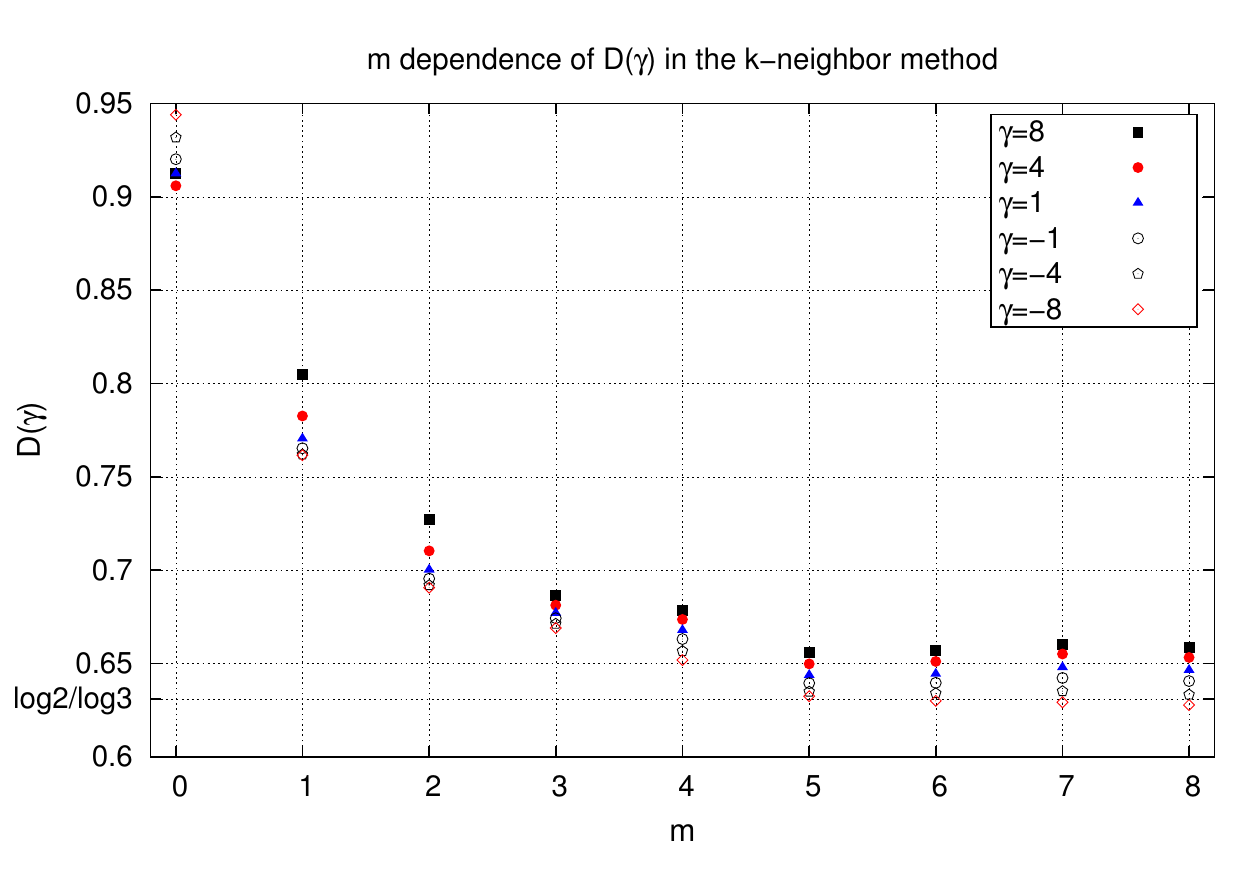}
\caption{These plots shows how the results for $D(\gamma)$ change as $m$ varies when the k-neighbor method is applied to the $m$\textsuperscript{th} finite representation of the uniform standard Cantor set. The theoretical value for $D(\gamma)$ is $\log (2)/ \log(3)$ for all $\gamma$. For all iterations the value of $n$ is fixed at 10000. The k-neighbor method provides relatively good results even when $m$ is as small as 5. }
\label{fig: k_small_m}
\end{figure}

 As a result of these findings, we have used two different boundaries for computing the slope, one for positive $\gamma$ and the other for negative $\gamma$, to produce the final results.  Since the inaccuracy inherited from these ambiguities cannot be entirely removed by increasing $n$ as in the nearest neighbor method, it is more difficult for the $k$-neighbor method to be adjusted to obtain a better result before knowing the theoretical values. Nevertheless, aside from these ambiguities in the method, the $k$-neighbor works for both positive and negative ranges of $q$, and therefore, is a good candidate as an initial method to investigate a given set. In the simulation, the ordering of the $n$ points from the reference points according to their relative position takes most of the computational time. Since the ordering takes more time as the topological dimension increases, the method is said to be especially suited for one-dimensional sets. 
Furthermore, unlike the nearest neighbor method, the hierarchy degree $m$ can be substantially small. The scaling region expectedly diminishes as $m$ decreases. However, the Dimension Function deduced from the best-linear-fit from the appropriate scaling region produces acceptable results. For the uniform Cantor Set, when $m$ is as small as 5, we obtained $D(\gamma)$ on the order of 0.6 as shown in Fig. ~\ref{fig: k_small_m}. This shows that to estimate the fractal dimension from the $k$-neighbor method, the finite representation does not necessarily require a large degree of hierarchy. Hence, the k-neighbor method is a good candidate for estimating the fractal dimensions when only a limited hierarchy degree is available. 
 
\section {Analysis \label{sec: analysis}}
\subsection{Range and Stability \label{subsec: range}}
In the nearest neighbor method, the probability distribution of $P(\delta, n)$ plays a key role as seen in Eq. \ref{eq: integral}. Hence, it is worthwhile to investigate the nature of probability distributions associated with fractal sets. Starting with the conjecture for the mathematical form for the cumulative distribution function for the uniform Cantor Set,
\begin{equation}
S(\delta,n)=1-\exp[-n(2\delta)^{D_0}] \label{eq: cumulative}
\end{equation}
Badii and Politi argue that the correct form of the probability density distribution of uniform Cantor set for $n>>1$ is given by \cite{Badii85} 
\begin{equation}
P(\delta, n) = 2 D_0 n (2 \delta)^{D_0 -1} \exp[-n(2\delta)^{D_0}] \label{eq: probability}
\end{equation}
Note that there is a singularity in the gamma function Eq. (\ref{eq: gamma_function}) for nonpositive integer $z$,\cite{abramowitz70} 
\begin{equation}
\Gamma(z) = \int_0^{\infty} t^{z-1}e^{-t} dt \label{eq: gamma_function}
\end{equation}
By substituting Eq. (\ref{eq: probability}) into (\ref{eq: integral}), a simple computation yields that 
\begin{align}\label{eq: singularity}
	M_\gamma(n) &= \left(\frac{1}{2n} \right) ^{\gamma/D_0}\int_{0}^{\infty} x^{\frac{\gamma}{D_0}} e^{-x} d x 
\\ &=   \left(\frac{1}{2n} \right) ^{\gamma/D_0} \Gamma(\gamma/D_0+1)
\end{align}
where $x=n(2 \delta)^{D_0}$. Therefore, the function, $M_\gamma(n)$, involves singularities for $\gamma < -D_0$. This means that, for the generalized Cantor set, the nearest neighbor is ill-suited for obtaining Correlation Dimension ($q=2$) or larger $q$. The result of $D(\gamma)$ for four different data sets are obtained using the nearest neighbor method as shown in Fig. ~\ref{fig: Typical}. In each plot, the numerical results are compared to the corresponding analytical results. The influence of the singularity is observed for a variety of sets. Note that the k-neighbor method does not suffer from this kind of singularity. For the $k$-neighbor method, the corresponding singularity can be found in Eq. (\ref{eq: k-3}). However, this time, the singularity can be avoided by taking a sufficiently large $k$. Accordingly, the $k$-neighbor method could generate sensible results in the entire range of $\gamma$ we have investigated. 

\begin{figure}[h]   
\includegraphics[scale=0.65]{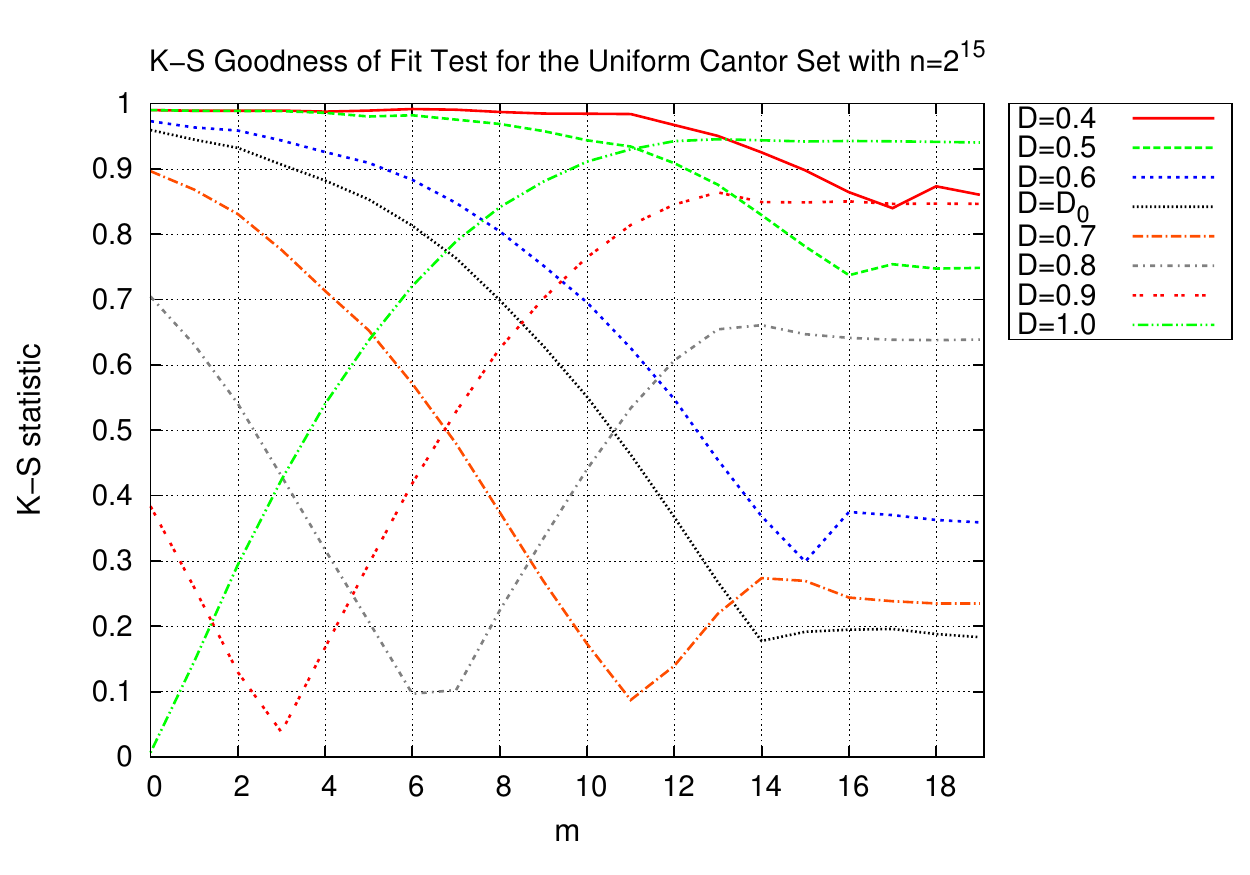}
\caption{\small \sl The Kolmogorov-Smirnov goodness-of-fit test was used to compare the simulated probability density distribution and the theoretical distribution proposed by Badii and Politi for the uniform Cantor set with $n=2^{15}$. According to Eq. \ref{eq: cumulative}, various values between $0$ and $1$ were substituted for $D_0$ for the purpose of this test.  Smaller values of the outcome indicate a better fit. The finite representation of the Cantor set with $m=1$ is the unit interval. Therefore, expectedly, the test function with $D=1$ exhibits the best fit among others. As $m$ increases, the K-S statistic decreases for $D=D_0={\ln2}/{\ln3}$ and similar values. However, they reach plateaus after $m=15$.  \label{fig: KS_test}}   
\end{figure}

It is worth noting that the simulated probability distribution functions did not completely converge to the theoretical distribution of Eq. (\ref{eq: probability}). The Komologov-Smirnov goodness-of-fit test measures the maximum discrepancy between two sample cumulative distributions and was employed to compare the theoretical distribution given by Eq. (\ref{eq: cumulative}) with $D_0=\frac{\ln2}{\ln3}$ and the distribution obtained in simulations. As seen in Fig. ~\ref{fig: KS_test}, the simulated distribution for the uniform Cantor set approaches the theoretical distribution when $D_0=\frac{\ln2}{\ln3}$ as $m$ increases. One would rationally expect the convergence to improve but this was not observed. When the number of intervals $2^{m}$ exceeds the number of points $N=2^{k}$, the nearest point for each reference point is likely to fall in the same interval, and therefore, the result of the K-S goodness-of-fit test constantly decreases when $m<k$. However, the maximum discrepancy reaches a plateau when $m=k$, suggesting that there is a constant disparity between the two distributions which does not diminish even when the finite representation of the Cantor set has large $m$ hierarchy degree. The results of the K-S test is shown in Fig. ~\ref{fig: KS_test} when the simulated distribution is compared against the theoretical distribution Eq. (\ref{eq: probability}) with different values for $D_0$. Among the values used, the theoretical distribution with $D=D_0=\ln2/\ln3$ showed the best fit for $m>14$. 
 
\begin{figure} [h] 
\begin{center}  
\includegraphics[scale=0.65]{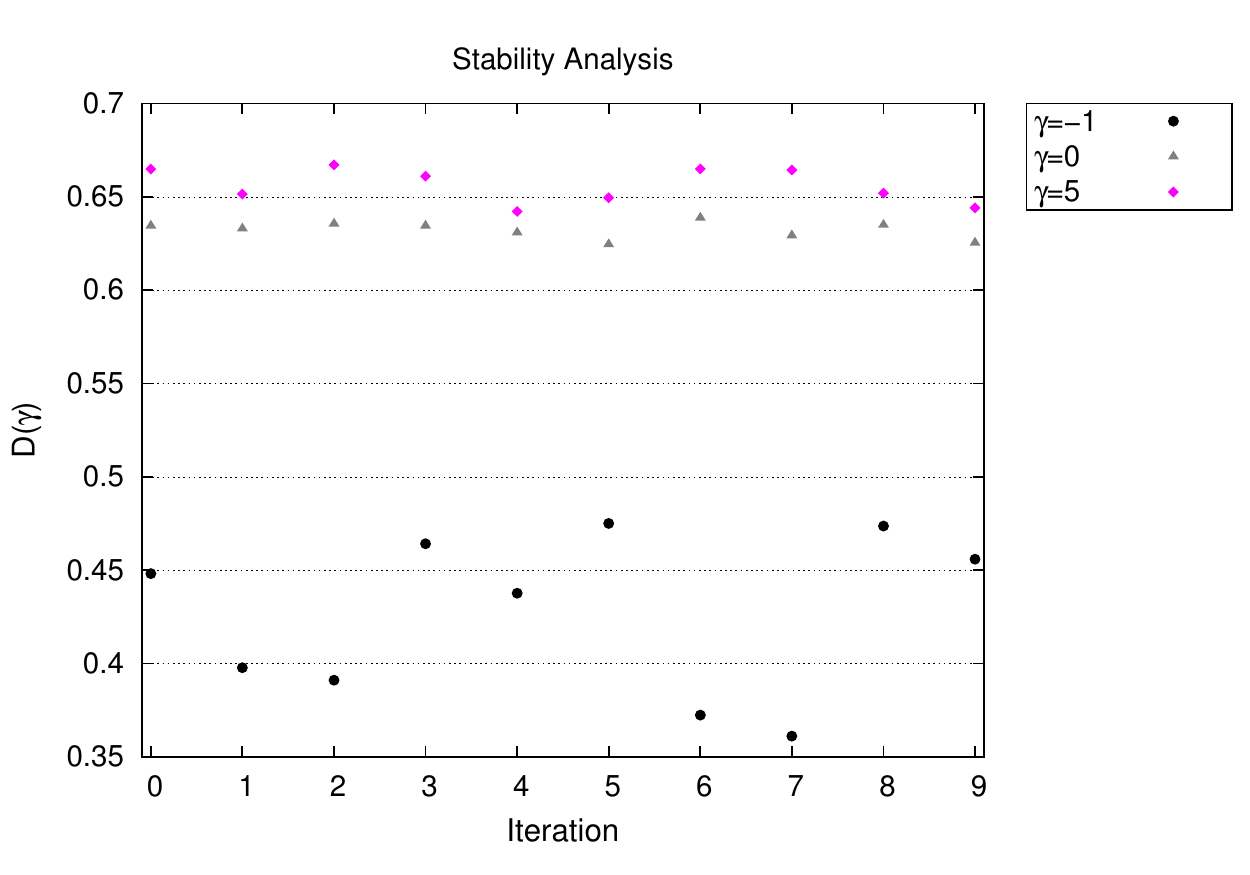}
\caption{\small \sl This figure shows that each iteration of the simulation generates a different outcome for$D(\gamma)$. Sample sets were taken from the uniform Cantor set. As $\gamma$ increases, the results fluctuate more. Larger fluctuation indicates more sensitive dependence on a particular choice of a sample set.  For negative $\gamma$, the outcome fluctuates even more and the average of the outcome is significantly smaller than the theoretical prediction which is roughly $0.63$.   \label{fig: Stability}}  
\end{center}  
\end{figure}

The effective domain is also related to the stability of the method. For both methods, as $|\gamma|$ increases, the nearest distance, $\delta$, is either amplified or attenuated. Consequently, the contribution from only a few sample points among $n$ chosen points starts to dominate the integral or sum in the equations. Unlike the nearest neighbor method, however, the effect of a few sample points is relatively small in the k-neighbor method due to the global feature. For the nearest neighbor method, simulations require a large number of ensembles and therefore, an extensive amount of computational time and memory for a relatively large negative $|\gamma|$.  How the Dimension Function $D(\gamma)$ varies in each implementation in the nearest neighbor method is shown in Fig. ~\ref{fig: Stability}. As $\gamma$ increases, the values of $D(\gamma)$ fluctuate more when computed under the same number of sample points. 

\begin{figure}[h]  
\begin{center}  
\includegraphics[scale=0.65]{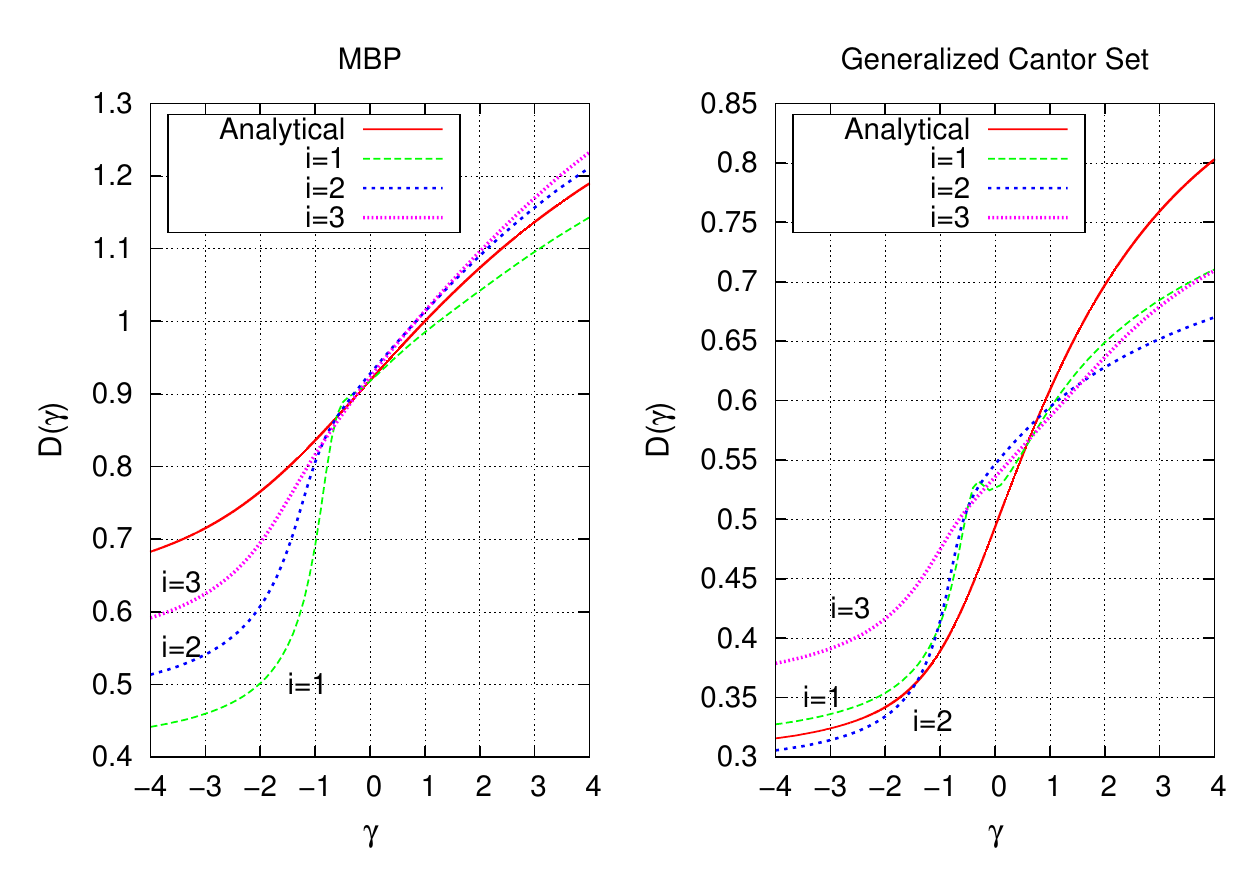}
\caption{\small \sl These plots show how using near neighbor instead the nearest neighbor affects the result. The integer $i$ denotes the $i$\textsuperscript{th} neighbor. While increasing $i$ generally makes $D(\gamma)$ smoother, one cannot expect that the results improve when $i$ is increased.  \label{fig: Near_neighbor}}  
\end{center}  
\end{figure}

This difficulty can be partially overcome by employing the ``near" neighbor instead of the nearest neighbor 
as it makes the simulation less dependent on the local property of a single reference point.  However, it eventually suffers from the same difficulty as the magnitude of $\gamma$ increases. The results for $D(\gamma)$ is shown in Fig. ~\ref{fig: Near_neighbor} when the near neighbor method is used. The integer $i$ denotes the $i$\textsuperscript{th}  neighbor points included in the partitions with $i=1$ being the nearest neighbor method. Moreover, as $i$ increases, all the relevant equations need to be modified accordingly but the dependence on $i$ is not obvious. Overall, the k-neighbor method has an advantage for large $|\gamma|$.

\subsection{The Limitation of Numerical Methods}
\begin{figure} [h] 
\begin{center}  
\includegraphics[scale=0.65]{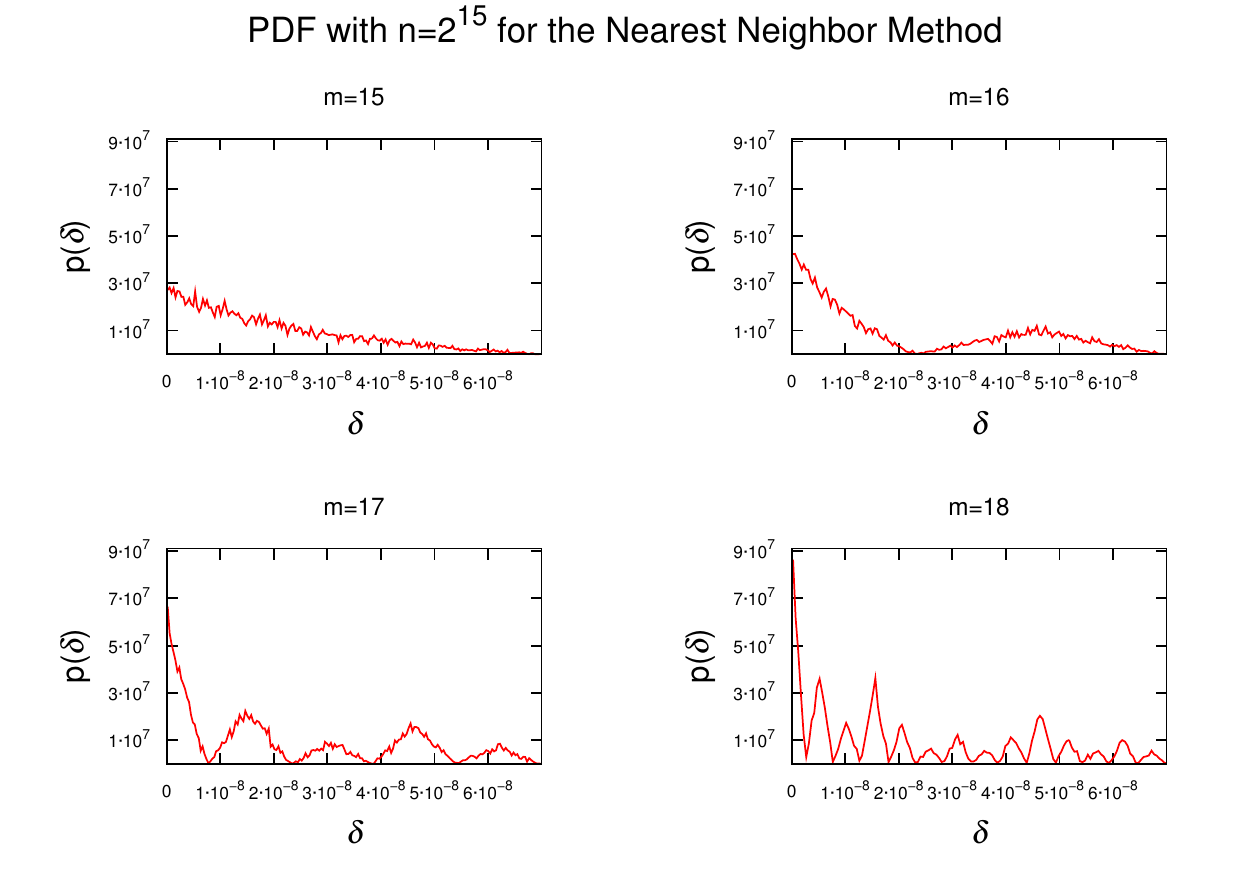}
\caption{\small \sl These plots show how the hierarchy degree $m$ affects the probability distribution of the nearest neighbor method. The sample sets were taken from the uniform Cantor set. While the cumulative distribution is somewhat more stable, as $m$ increases, the fine structure of the probability distribution of $\delta$ emerges, exhibiting self-similar patterns. A limited horizontal range from $0$ to $3^{-15}$ is plotted. \label{fig: Fine}}  
\end{center}  
\end{figure}

\begin{figure}[h]  
\begin{center}  
\includegraphics[scale=0.65]{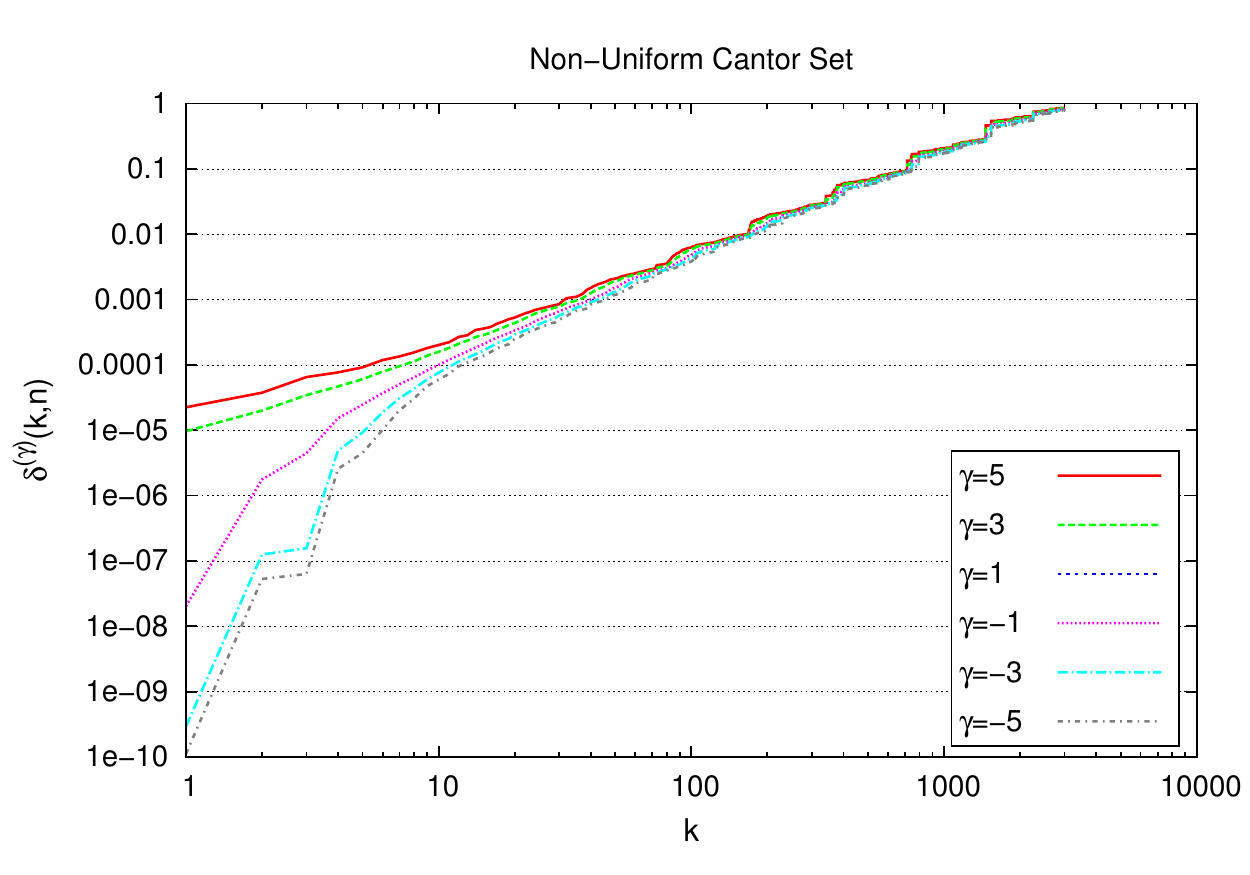}
\caption{\small \sl $\delta^{\gamma}(k,n) =  \left(\Delta^{\gamma}(k,n) \right)^{(1/\gamma)}$ is plotted versus $k$ in a log-log plot. The fine structure inherited from the non-uniform Cantor set is observed. \label{fig: K}}  
\end{center}  
\end{figure}

As shown in Fig. ~\ref{fig: Fine} and \ref{fig: K}, plots of the probability distribution $P(\delta, n)$ of $\delta$ for the nearest neighbor method or the $k$\textsuperscript{th} neighbor distance $\delta^{\gamma}(k,n)$ typically exhibit self-similar fine structures which arise from the original fractal geometry. However, unless a construction recipe is known in advance, as in the case of the generalized Cantor set, the exact nature of the fine structure is difficult to obtain. Moreover, to find its exact nature is essentially redundant for it would be another fractal set which is as complex as the original fractal set. Hence, numerical methods are typically developed based on an assumption that these fine structures will not affect their outputs in any substantial way. Nevertheless, we should not simply ignore the effect of the fine structures as a set would not be a fractal without them. In the equations such as Eqs. (\ref{eq: integral}) and (\ref{eq: k-2}), the fine structures are absorbed by the constant or correction term. In general, these correction terms depend on the hierarchy degree used in creating a test set as well as the number of sample points. However, it is difficult to estimate the error attributed to the correction term, and therefore this raises a question concerning the reliability of the method.  

In principle, the largest possible $m$ should be used to reflect the infinite hierarchical self-similarity. For the nearest neighbor method, the number of reference points, $n$, needs to be smaller than $2^m$. Therefore, to increase $n$ to obtain more accurate results, one needs to increase $m$ as well. However, unlike in the case of sample points where increasing $n$ generally guarantees a more accurate result, increasing $m$ does not necessarily. As you can see in Fig. ~\ref{fig: k_small_m}, once $m$ reaches a certain threshold, increasing $m$ will not produce a better result. 

\section{Conclusion \label{sec: conclusion}}
In contrast with the box-counting method, or similar methods which utilize partitions of equal sizes, we have shown that the nearest neighbor method, which employs partitions of equal mass, as well as the $k$-neighbor method, which employs partitions of  distributed mass, are good candidates for estimating the generalized fractal dimension for negative $q$. The $k$-neighbor method works for the complete range of $q$ and no serious deviations were found. By choosing an appropriate scaling region, it is possible to estimate the generalized dimensions even with a small hierarchy degree. However, the method involves linear regression and the results depend on how the best-fit line is obtained. Therefore, the $k$-neighbor method is a good option for a starting point and to investigate the general outlook of $D_q$. If the sample size is large, the nearest neighbor method can be the best method for small negative $q$. Although the result is sensitive to the local anomalies, one can choose the size of $n$ according to one's  required precision to extract the dimension. However, in contrast with the $k$-neighbor method, the hierarchy degree, $m$, also needs to be sufficiently large in order to obtain a desirable probability distribution. Therefore, if the sample size of a finite representation is small, the nearest neighbor method is not a practical choice. For positive $q$, the methods with partitions of equal sizes may be used. In general, a few different methods should be applied before one determines if the results from different methods are consistent. The $k$-neighbor method should  provide the overall features of $D_q$.  Given that the subjective choice of the best-fit line affects the result, it is important to determine the window of ambiguity. If the sample size is adequate, apply the nearest-neighbor method for negative $q$ and box-counting or similar method for positive $q$. The results from these two different methods should lie within the window of ambiguity. 

	In any simulation of the kind worked out in this paper, the finite sample correction needs to be taken care of. Although a number of correction terms have been proposed over the years, \cite{Water88, broggi88} many of them add extra complications to the simulation without achieving a dramatic increase in their method's accuracy. \cite{grassberger1985generalizations, broggi1988evaluation, Water88} In the process of exploring the form of the nearest neighbor distribution of the generalized Cantor set, some interesting properties have been obtained; the order of taking $m$ and $n$ to infinity may not commute as usually assumed. Since a numerical sample only possesses a finite hierarchy, a new algorithm which does not assume an infinite hierarchy may be useful. In future work it will be shown that a new analysis of generalized dimension may be based on some quantities that are independent of the hierarchy. 

% If in two-column mode, this environment will change to single-column format so that long equations can be displayed. 
%\begin{widetext}
%$$\mbox{put long equation here}$$
%\end{widetext}

% Figures should be put into the text as floats. 
% Use the graphics or graphicx packages (distributed with LaTeX2e).
% Put the label that you will use with \ref{} command in the braces of the \label{} command.
%
% \begin{figure}
% \end{figure}

% \begin{table}
% \end{tabular}
% \end{table}

\begin{acknowledgments}
We would like to thank Dr. Igor Prokhorenkov for his insightful and valuable advice. 
\end{acknowledgments}


\end{document}